\documentclass[useAMS,usenatbib]{mn2e}
\usepackage{graphicx,natbib}

\begin{document}

%
%

\newif\ifAMStwofonts
\AMStwofontstrue


\newcommand{\kms}{~km~s$^{-1}$}
\newcommand{\vrot}{$v_{\rm rot}$}
\newcommand{\mdot}{$\rm{M}_{\odot}$}
\newcommand{\chisq}{$\chi^{2}$}
\newcommand{\vsini}{$v\sin i$}

\def\aap{Astron.~Astrophys.~}
\def\aaas{Astron.~Astrophys.~Suppl.~}
\def\mnras{Mon.~Not.~R.~Astron.~Soc.~}
\def\aas{Astron.~Astrophys.~}
\def\aas{Astron.~Astrophys.~Suppl.~}
\def\aj{Astron.~J.~}
\def\apj{Astrophys.~J.~}
\def\apjl{Astrophys.~J.~Lett.~}
\def\apjs{Astrophys.~J.~Suppl.~}
\def\baas{Bull.~Am.~Astron.~Soc.~}
\def\pasp{Publ.~Astron.~Soc.~Pac.~}
\def\pasj{Publ.~Astron.~Soc.~Jap.~}
\def\araa{Ann.~Rev.~Astron.~Astrophys.~}
\def\an{Astron.~Nachr.~}
\def\apss{Astrophys.~Sp.~Sci.~}

\title{Doppler Images and Chromospheric Variability of TWA 6}
\author[M.~B.~Skelly et al]{M.~B.~Skelly,$^{1}$\thanks{E-mail: mairead.skelly@imperial.ac.uk}  
Y.~C.~Unruh,$^{1}$ A. Collier Cameron,$^{2}$ J. R. Barnes,$^{3}$ J.-F. Donati,$^{4}$ 
\newauthor
 W. A. Lawson,$^{5}$ B. D. Carter$^{6}$ \\
$^{1}$ Astrophysics Group, Imperial College of Science, Technology and Medicine, London SW7 2AZ \\
$^{2}$ School of Physics and Astronomy, University of St Andrews, Fife KY16 9SS \\
$^{3}$ Centre for Astrophysics Research, University of Hertfordshire, College Lane, Hatfield, Herts AL10 9AB  \\
$^{4}$ Observatoire Midi-Pyr\'{e}n\'{e}es, LATT, 14 avenue Edouard Belin, 31400 Toulouse, France \\
$^{5}$ School of Physical, Environmental and Mathematical Sciences, University of New South Wales, \\
Australian Defence Force Academy, Canberra, ACT 2600, Australia \\
$^{6}$ Faculty of Sciences, University of Southern Queensland, Toowoomba, 4350, Australia }

\date{3 January 2008}

\pagerange{\pageref{firstpage}--\pageref{lastpage}}
\pubyear{2007}

\newtheorem{theorem}{Theorum}[section]

\label{firstpage}
\maketitle

\begin{abstract}
We present Doppler imaging and Balmer line analysis of the weak-line T Tauri star TWA 6. Using this data we have made one of the first attempts to measure differential rotation in a T Tauri star, and the first detection of a slingshot prominence in such a star. We also show the most direct evidence to date of the existence of solar-type plages in a star other than the Sun. 

Observations were made over six nights: 11--13th February 2006 and 18--20th February 2006, when spectra were taken with the UCL Echelle Spectrograph on the 3.9-m Anglo--Australian Telescope. Using least-squares deconvolution to improve the effective signal--to--noise ratio we produced two Doppler maps. These show similar features to maps of other rapidly rotating T Tauri stars, i.e. a polar spot with more spots extending out of it down to equator. Comparison of the two maps was carried out to measure the differential rotation. Cross-correlation and parameter fitting indicates that TWA 6 does not have detectable differential rotation.  

The Balmer emission of the star was studied. The mean H$\rm \alpha$ profile has a narrow component consistent with rotational broadening and a broad component extending out to $\pm$250\kms. The variability in H$\rm \alpha$ suggests that the chromosphere has active regions that are cospatial with the spots in the photosphere, similar to the `plages' observed on the Sun. In addition the star has at least one slingshot prominence $3R_{*}$ above the surface -- the first such detection in a T Tauri star.
\end{abstract}


\section{Introduction}

\subsection{T Tauri Stars}

T Tauri stars are pre-main sequence equivalents of Sun-like stars. They are young, have low masses (0.3 -- 1.0~\mdot) and spectral types F--M~\citep{hayashi66}. The T Tauri phase is considered to begin when the star becomes optically visible and continues until hydrogen burning begins and the star has reached the post-T Tauri phase at an age of the order of $10^{7}$ years (depending on spectral type). 

Very young stars fall on Hayashi tracks on the Hertzsprung-Russell diagram, i.e. vertical tracks of constant temperature. As the star approaches the zero-age main sequence (ZAMS) it can either continue to follow the Hayashi track all the way onto the ZAMS (and remain fully convective), or develop a radiative core and turn onto the Henyey track and evolve onto the ZAMS with constant luminosity and increasing temperature. Which of these evolutionary paths the T Tauri star follows is thought to depend on the mass of the star but observations of TTS with ages close to this turn-off will allow us to characterise this stage of stellar evolution more fully. 

T Tauri stars fall into two subcategories -- classical and weak-line (\citealp{bertout89} and references therein). Observations of classical TTS (CTTS) show excess ultraviolet and infrared emission and strong emission line activity, characteristics which are due to the presence of a circumstellar disc. The star--disc interactions lead to strong, irregular variability that is predominately due to accretion of material from the disc onto the surface. The disc has a braking effect on the star, causing CTTS to rotate more slowly than their discless equivalents \citep{bouvier93,cc93}. 

Weak-line T Tauri stars (WTTS) have weaker emission-line activity and show less evidence of accretion. WTTS appear to be older than CTTS on average \citep{armitage03},  suggesting that when TTS form they have circumstellar discs which dissipate as the star evolves, but disc lifetimes are variable, ranging from 0.1 -- 10~Myr \citep{bouvier97}. WTTS are variable, but in contrast to CTTS the variation is usually periodic as it is due to cool spots that go in and out of view as the star rotates. These cool spots form when magnetic flux tubes emerge through the photosphere. 

Doppler imaging \citep{vogt83} maps spots on stellar photospheres. Doppler imaging of rapidly rotating, magnetically active stars, such as WTTS, often finds spots at high latitudes and even at the poles, see e.g. \cite{v41096}, \cite{strass02}, in contrast to the Sun where they are normally within $\pm30^\circ$ of the equator. One of the favoured explanations is that Coriolis forces in rapidly rotating stars cause the flux tubes to emerge at higher latitudes \citep{schu92,caligari94}. 

In the past Doppler imaging of cool stars has been limited by its challenging signal-to-noise requirements. The technique of least-squares deconvolution \citep[LSD,][]{jf97} has made it possible to produce Doppler images of stars for which the signal-to-noise ratio (SNR) would be too low if individual spectral lines were used. In LSD the underlying rotationally broadened line profile for each spectrum is found by deconvolving a line list from the spectrum. This makes it possible for late--K T Tauri stars to be imaged, where previously the SNR requirements would have been prohibitive. 

Sunspots are accompanied by prominences, i.e. loops of material that follow the magnetic field lines emerging through the surface of the star. High--lying prominences, known as `slingshot prominences', typically appear as fast--moving absorption transients and have been observed in a number of young stars such as Speedy Mic \citep{duns06} and AB Dor \citep{cc89,abdor99}, but not in any TTS to date. These prominences are viewed as a tracer for large-scale magnetic fields. If a T Tauri star was shown to be able to support such high-lying prominences for several rotational periods it would suggest the presence of a large-scale field that does not decay away quickly outside the stellar radius.

\subsection{Differential Rotation in Stars with Deep Convection Zones}\label{intdiff}

Fully convective stars are believed to have a distributed magnetic dynamo ($\alpha^{2}$), arising from turbulent motions or Coriolis forces, rather than a solar-type dynamo ($\alpha\Omega$) that is due to both convection and differential rotation, as seen in stars with radiative cores. 

Differential rotation, when rotational velocity is a function of latitude, is expected to be less apparent with increasing convection zone depth \citep{kitch99}, so stars with larger convection zones are expected to rotate more like a solid body. Differential rotation in the Sun gives rise to the $\Omega$ effect that produces a large-scale toroidal field from the small-scale poloidal field. Most models for an $\alpha^{2}$ dynamo rule out the possibility of there being differential rotation in fully convective stars \citep[e.g.][]{chab06}. However \cite{dobler06} suggest fully convective stars have significant differential rotation, although this would not be Sun-like, and takes the form of a meridional circulation. 
 
Observations to date have only attempted to detect the latitudinal differential rotation seen in the Sun. A recent measurement for the fully-convective M4 dwarf V374 Peg \citep{morin07} found a weak differential rotation (around 0.1 of the solar value). Differential rotation has been observed to fall with decreasing photospheric temperature. For example, figure 2 in \cite{barnes05a} plots differential rotation against temperature. The trend seems to imply that differential rotation approaches zero below 4000K but it is uncertain where it becomes undetectable as there has been no measurement to date of stars with temperatures between 3800--4500K. In addition there have not been any reliable measurements of differential rotation in T Tauri stars. Such measurements are needed to confirm that TTS follow these trends too.
\subsection{TW Hydrae Association}

Nearby young co-moving groups are a rich source of T Tauri stars for observation. Our chosen targets lie in the TW Hydrae association (TWA), a group of pre-main sequence stars lying at a distance of about $51\pm 5$pc and with an age of 8 Myr \citep{zuck04}. Of the 24 likely members of the TWA, 16 have had their periods measured in \cite{lawson05}. The age of this association is an important one as it is corresponds to the mean age of circumstellar discs and hence the time when T Tauri stars can begin to spin up, so the members of this association are interesting objects for study.
 The star that gives the association its name (TW Hya) is a classical T Tauri star, spectral type K7. Its disc has been directly imaged in \cite{twdisk} and its magnetic field was detected in \cite{yang07}. Although it is rotating fairly rapidly (P = 2.8d) it is almost pole--on and is therefore not a suitable candidate for Doppler imaging.

This paper concerns another member of the TW Hydrae association -- TWA 6. It has a spectral type of K7 \citep{webb99}, corresponding to a temperature of about 4000K for a T Tauri star \citep{cohen79}. Models by \cite{siess00} indicate that it is fully convective or just beginning to develop a radiative core. Photometry has revealed a large lightcurve amplitude, with a variation of 0.49 magnitudes in the V band \citep{lawson05}. Its rapid rotation, \vsini~ = 55~\kms~ \citep{webb99} and period of 0.54 days \citep{lawson05} make it an excellent target for Doppler imaging. In this paper we present Doppler images of TWA 6, and a discussion of whether it has differential rotation. We also analyse the Balmer line emission in order to study the behaviour of the chromosphere. 

\section{Observations and Data Reduction}

Observations were carried out at the Anglo-Australian Telescope (AAT) using the University College London Echelle Spectrograph (UCLES) and the EEV2 $2\rm{k} \times 4\rm{k}$ detector between 11--20th February 2006. In order to detect differential rotation observations were made with a gap of 4 nights. By doing this two maps can be made of the star allowing differential rotation to be detected, or an upper limit placed on its value \citep{petit04}. With three nights of observations on either side of this gap full phase coverage was achievable on these two separate occasions.

We took spectra of the targets, telluric and spectral standard stars, and calibration frames (see tables \ref{obs1} and \ref{obs2}). UCLES was centred at $5500\rm{\AA}$ giving a wavelength coverage of $4400 - 7200\rm{\AA}$. A slit width of 1.0\arcsec~was used giving a spectral resolution of 45000.

The weather on the first three nights was reasonably good, and although conditions and seeing were variable (seeing ranged from 1.2--2.2\arcsec) in total only 3~hrs were lost. 36 good quality spectra of TWA 6 were obtained, as well as 27 spectra of our secondary target TWA 17 (Skelly et al, in prep.). On the second set of nights the weather was of variable quality and more time was lost. Of a potential 24 hours we had 13 hours of observing time, with seeing between 1.3 -- 2.2\arcsec. Most of this time were concentrated on TWA 6 and a further 39 spectra were obtained. An additional 1.5hr of observing time was obtained on 17th Feb 2006 due to the early finish of aluminising the telescope mirror. 

\begin{table*}
\begin{minipage}{120mm}
\centering
\caption{Table setting out the observations made on the nights of 11/02/06
to  13/02/06. Spectral types in the `Comments' column refer to spectral standards. The signal-to-noise ratio is the maximum for the set of exposures.}
\begin{tabular}{lrcrcrll}
\hline

Object &  Date &  UT Start & Exp. & No. of & S/N & Comments & Conditions \\
& & & time [s] & Frames & & & \\

\hline

HR 3037 & 11 & 09:42:56 & 20 & 1 & 98 & Telluric & \\
HR 3037 & 11 & 09:46:15 & 120 &  1 & 164 & Telluric & \\
HD 34673 & 11 & 09:58:05 & 200 & 1 & 68 & K3  & \\
TWA 6 & 11 & 10:25:43 & 900 & 6 & 53 & Target 1 & \\
HR 3037 & 11 & 12:10:19 & 20 & 1 & 105 & Telluric &  \\
HD 52919 & 11 & 12:16:17 & 200 & 1 & 79 & K5 &  \\
TWA 6 & 11 & 12:22:44 & 900 & 5 & 63 & Target 1 &  \\
SAO 116640 & 11 & 13:45:02 & 450 & 1 & 110 & K7 &  \\
TWA 6 & 11 & 13:55:31 & 900 & 5 & 58 & Target 1 &  \\
TWA 6 &  11 & 15:17:37 & 900 & 5 & 62 & Target  1 &  \\
TWA 17 & 11 & 16:43:21 & 1200 & 1 & 48 & Target 2 &  \\
TWA 6 & 11 & 18:01:59 & 900 & 4 & 53 & Target  1 &  \\
\hline
GJ 204 & 12 & 09:40:18 & 120 & 2 & 58 & K5 &  \\
GJ 3331A & 12 & 10:01:49 & 900 & 1 & 16 & M2 & Diffuse cloud \\
TWA 6 & 12 & 10:23:33 & 900 & 1 & 19 & Target 1 & Diffuse cloud \\
TWA 6 & 12 & 10:42:47 & 900 & 1 & 3 & Target  1 & Aborted 148s \\
HD 19007 & 12 & 11:04:33 & 200 & 1 & 40 & K4  & Heavy cloud \\
TWA 17 & 12 & 12:42:06 & 1200 & 1 & 31  & Target 2  & \\
HR 3037 & 12 & 13:06:28 & 20 & 1 & 90 & Telluric & \\
TWA 17 & 12 & 13:13:37 & 1200 & 2 & 33 & Target 2  &  \\
TWA 17 & 12 & 14:06:49 & 1200 & 3 & 42 & Target 2 &  \\
HR 3037 &  12 & 15:26:56 & 20 & 1 & 40 & Telluric & Cloud  \\
TWA 17 & 12 & 15:42:30 & 1200 & 2 & 28 & Target  2 & Cloud  \\
HR 3037 & 12 & 16:48:51 & 20 & 1 & 87 & Telluric &  \\
TWA 17 &  12 & 16:55:16 & 1200 & 3 & 36 & Target  2 & Some fog  \\
\hline
HR 3037 & 13 & 09:32:42 & 20 & 1 & 49 & Telluric &  \\
HD 52919 &  13 & 09:35:50 & 200 & 1 & 90 & K5 &  \\
SAO 116640 &  13 & 09:40:57 & 400 & 1 & 82 & K7 &  \\
GJ 3331A &  13 & 09:50:39 & 900 & 1 & 18 & M2 &  \\
TWA 6 &  13 & 10:09:24 & 900 & 9 & 37 & Target 1 & Seeing $>$ 2'' \\
GJ 204 & 13 & 12:36:26 & 120 & 1 & 67 & K5 &  \\
TWA 17 & 13 & 12:44:30 & 1200 & 2 & 23 & Target 2 & Seeing $>$ 2'' \\
TWA 6 &  13 & 13:31:51 & 900 & 1 & 39 & Target 1 &  Seeing $>$ 2''  \\
TWA 17 & 13 & 13:49:34 & 1200 & 2 & 30 & Target 2 & Seeing $>$ 2'' \\
TWA 6 & 13 & 14:33:20 & 900 & 1 & 42 & Target 1 & Seeing $>$ 2'' \\
TWA 17 &  13 & 15:01:47 & 1200 & 2 & 27 & Target 2 &  \\
TWA 6 & 13 & 15:35:40 & 900 & 1 & 44 & Target 1 &  \\
TWA 17 & 13  & 15:54:08 & 1200 & 4 & 31 & Target 2 &  \\
TWA 6 & 13 & 17:22:43 & 900 & 1 & 49 & Target 1 &  \\
TWA 17 & 13  & 17:40:59 & 1200 & 4 & 34 & Target 2 &  \\

\hline
\end{tabular}
\label{obs1}
\end{minipage}
\end{table*}

\begin{table*}
\begin{minipage}{120mm}
\centering
\caption{As table \ref{obs1} for observations made on the nights of 17/02/06 to 20/02/06.}
\begin{tabular}[8pt]{lrcccrll}
\hline

Object &  Date &  UT Start & Exp. & No. of & S/N & Comments & Conditions \\
& & & time[s] & Frames & & & \\

\hline
HR 3037 & 17 & 17:07:40 & 20 & 1 & 75 & Telluric & Late start \\
TWA 6 &  17 & 17:10:02 & 900 & 6 & 59 & Target &   \\

\hline

GJ 529 & 18 & 16:53:38 & 200 & 1 & 38 & K4.5 & Storms  \\
TWA 6 & 18 & 17:14:09 & 900 & 4 & 67 & Target \\
BD -20 4645 & 18 & 18:23:37 & 650 & 1 & 114 & K6  & \\

\hline

HR 3037 &  19 & 09:55:30 & 20 & 1 & 121 & Telluric & Cloud/seeing $> 2''$ \\
BD -02 801 & 19 & 10:02:10 & 800 & 1 & 115 & K6  & Cloud/seeing $>2''$ \\
TWA 6 &  19 & 10:18:29 & 900 & 3 & 42 & Target 1 & 3rd aborted 768s \\
HR 3037 &  19 & 15:05:55 & 20 & 1 & 74 & Telluric &  Storm  \\
TWA 6 & 19 & 15:08:40 & 900 & 8 & 58 & Target 1 & Some cloud  \\
TWA 17 &  19 & 17:19:03 & 1200 & 1 & 46 & Target 2 & Some cloud  \\
TWA 6 & 19 & 17:41:09 & 900 & 4 & 54 & Target 1 & Some cloud    \\

\hline

HR 3037 & 20 & 09:40:29 & 20 & 1 & 110 & Telluric & Cirrus cloud \\
TWA 6 & 20 & 09:45:46 & 900 & 14 & 38 & Target 1 & Cirrus cloud \\
HR 3037 & 20 & 14:36:28 & 20 & 1 & 97 & Telluric & Cirrus cloud   \\
TWA 6 & 20 & 14:39:22 & 900 & 10 & 51 & Target 1 & Cirrus cloud  \\
TWA 17 & 20 & 17:22:40 & 1200 & 2 & 31 & Target 2 & Cirrus cloud  \\

\hline
\end{tabular}
\label{obs2}
\end{minipage}
\end{table*}

\subsection{Data Reduction}

Data reduction was carried out with the {\sc STARLINK} routine {\sc ECHOMOP} \citep{echomop1997}. Frames were bias subtracted and flatfielded, cosmic rays were removed and the spectra were extracted. Sky subtraction was carried out by subtracting a polynomial fitted to the inter-order gaps. The spectra were normalised using a continuum fit to a star of the same spectral type. The spectral type (K7) was verified by eye by identifying the best temperature fit to the spectra in comparison with the template spectra. Wavelength calibration was carried out using ThAr arc spectra. 

\section{Least-squares Deconvolution}

As the 36+39 spectra did not have a high enough SNR for Doppler imaging we used least-squares deconvolution \citep[LSD, described in][]{jf97} to extract the underlying rotationally broadened profiles. LSD uses matrix inversion to deconvolve a line list from a spectrum to leave a high signal-to-noise profile. The program {\sc SPDECON} \citep[implemented as described in][]{barnes98} was used to do this. We used a line list from the Vienna Atomic Line Database \citep[VALD,][]{vald} for a 4000K star. Over 4000 images of 2000 spectral lines were used in the deconvolution leading to a signal gain of $\sim 60$. For the deconvolution, the lines were weighted by $\rm{counts}^{2}/\rm{variance}$. The effect of this is to favour regions of the spectrum with higher SNR, and the centres of the orders rather than the wings. The resultant line profiles had a mean centroidal wavelength of $5530 \rm \AA$. 

\begin{figure}
\includegraphics[width=0.45\textwidth]{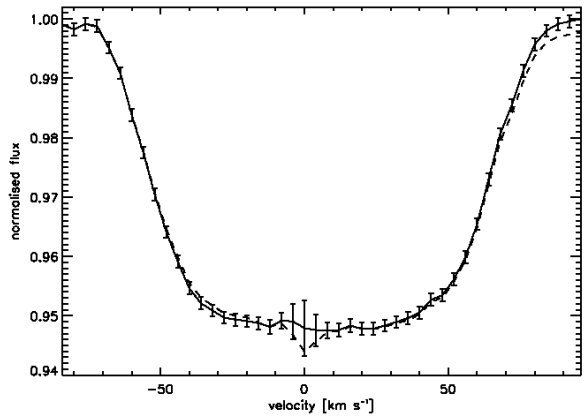}
\caption{The dashed line shows the mean deconvolved profile for the first
dataset (11--13th Feb 2006). There is an absorption feature due to sunlight reflected off the Moon at 0~\kms.
The solid line is the mean of the profiles after this feature was removed
and a slope was removed. Mean errors are shown for the data after the feature was removed. The errors close to 0 \kms have been increased due to the subtraction.}
\label{moon}
\end{figure}

There was a small consistent tilt in the profiles which was removed by fitting a straight line between the two sections of continuum on the left and right sides. This tilt changed depending on which spectral standard was used and hence is most likely due to a small mismatch between the temperatures of the star and the spectral standards. In addition there was a feature near 0\kms~in many of the profiles. It was identified as being due to moonlight in the spectra as it only appeared to be present when the Moon was above the horizon (Fig \ref{moon}).

As we did not have a solar spectrum we deconvolved the hottest template spectrum (K3) we had taken with UCLES to model the shape of this feature. We were unable to find a pattern to its depth as it is difficult to disentangle from variations due to surface features. We failed to find a relationship between its equivalent width and the zenith distance, time or seeing, although the absorption signal was stronger during bad weather. Hence for all spectra taken when the Moon was up, a constant shape was subtracted from all profiles taken within the same night (see Fig. \ref{moon}). This was calculated by inspecting the mean of all profiles, and subtracting a profile such that the absorption feature was not obvious in the mean. 
A scale factor was calculated separately for each night. The errors for the pixels affected were increased so that these pixels would have a lower influence on the final image produced. This was done by adding in quadrature the error produced by {\sc SPDECON} and the amount that was subtracted from each pixel.

\begin{figure}
\includegraphics[width=0.48\textwidth]{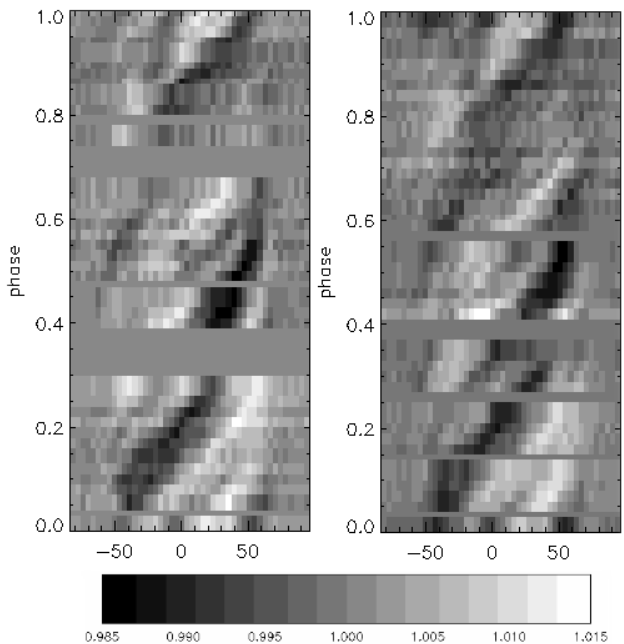}
\caption{Greyscale images of the LSD profiles divided by the mean profile for 11--13th Feb (left) and 17--20th Feb (right). Spots appear as emission (white). The greyscale is given in the bar at the bottom.}
\label{grey}
\end{figure}

Each of the deconvolved profiles has been normalised by dividing through by the mean profile. In Fig. \ref{grey} the normalised profiles have been stacked and shown in greyscale. As spots appear in emission the lighter regions in these images show the spots being carried across the stellar disc. Spot groups are apparent at the same phases in both, particularly around phases 0.05 and 0.55, corresponding to longitudes of $340^{\circ}$ and $160^{\circ}$. This suggests that the spot distribution is similar at both epochs. 
\section{Doppler Imaging}
\subsection{Doppler Imaging Code}

The unnormalised profiles and corresponding errors produced by {\sc SPDECON} are used as inputs for the program DoTS \citep{dots01}. The program calculates the surface temperature distribution by using \chisq~minimization to find the best fit to the data. A two-temperature model is used to describe the star \citep{cc94} and each element of the star is assigned a spot-filling factor $f$. DoTS finds an optimal solution by minimising \chisq~while maximising the entropy of the image. It is not sufficient to simply minimize \chisq~as a large number of images may fit the data, particularly when the errors are large. The maximum entropy approach allows a unique image to be returned that does not contain information that was not required by the data \citep{maxent}. DoTS also requires a look-up table, containing the shape of the unbroadened spectral line at a number of limb angles and at the spot and photospheric temperatures. 

\subsection{Stellar Parameters}\label{parsec}

Many of the parameters of TWA 6 were known from previous work and are listed in table \ref{param2}. Also listed are our new or updated parameters as determined using $\chi^{2}$ minimization. Testing the program with artificial data has shown that nearly all of the parameters can be derived independently by finding the value which gives the lowest $\chi^{2}$ (even if the other parameters have not yet been fixed). However, for some parameters the minimization must be carried out for two parameters simultaneously, e.g. period and differential rotation must be determined together (see section \ref{diff} for further discussion of differential rotation). Of the previously measured values only one changed by a notable amount: the projected rotational velocity, \vsini~was found to have a value of 72\kms, somewhat higher than the previously measured value of 55\kms~\citep{webb99}. 

Constraining the inclination is slightly more difficult. The minimum of the \chisq~plot is not as clear as with some of the other parameters and the curve is very shallow around the minimum \chisq, found at an inclination of $47^{\circ}$. This difficulty in obtaining the inclination angle has been noted in other work e.g. \cite{kurster94} where $\pm{10^{\circ}}$ is given as a reasonable range. The errors given in table \ref{param2} were determined by eye from the width of the `trough' in the \chisq~vs $\sin i$ plot.

Using the inclination angle, period and projected equatorial velocity the radius was calculated. The resulting value of $1.05R_{\odot}$ is within one sigma of the value calculated using the distance and V - R colour of $(0.8\pm0.4)R_{\odot}$, using the method described in \cite{gray92}. It should be noted that the distance quoted in table \ref{param2} is determined using Hipparcos distances to other members of the TWA and is therefore uncertain so the radius determination using the rotational velocity and period is the more reliable one. 

Assuming an age of 8 Myr, an effective temperature of 4000K and a luminosity of 0.25 $\rm{L}_{\odot}$ the best-fitting mass of the star, from both the \cite{alex89} and the \cite{kurucz91} opacities \citep{dantona94}, is 0.7\mdot. The luminosity was calculated using the radius and effective temperature.
\begin{table}
\centering
\caption{Previously-known and updated parameters for TWA 6. References: [1]: \protect\cite{zuck04}, [2]: \protect\cite{cohen79}, [3]: \protect\cite{webb99}, [4]:\protect \cite{lawson05}, [5]: \protect\citep{dantona94}.}
\begin{tabular*}{0.48\textwidth}{llll}
\hspace{-2mm}Parameter & Previous & New Value & Reference \\
& Value & & \\
\hline
\hspace{-2mm}Spectral type & K7 & K7 & [1] \\
\hspace{-2mm}Temp. [K] & -- & $4000\pm200$ & [2] \\
\hspace{-2mm}$V$ magnitude & 12  & -- & [3] \\
\hspace{-2mm}$V~-~R$ & 1.19 & -- & [2] \\  
\hspace{-2mm}Distance [pc] & $51\pm 5$ & -- & [1] \\
\hspace{-2mm}\vsini~[kms$^{-1}$] & $55\pm 10$ & $72\pm 1$ & [3], this work  \\
\hspace{-2mm}Period [day] & $0.54\pm0.01$ &$0.5409\pm0.00005$& [4], this work \\
\hspace{-2mm}Inclination [$^{\circ}$] & -- & $47^{+10}_{-8}$ & this work \\
\hspace{-2mm}Radius [$\rm{R}_{\odot}$] & -- & $1.05^{+0.16}_{-0.15}$ & this work \\ 
\hspace{-2mm}Mass [\mdot] & -- & $0.7\pm0.2$ & [5], this work \\
\hspace{-2mm}Lum. [$\rm{L}_{\odot}$] &$0.16^{+0.13}_{-0.08}$ & $0.25\pm{0.04}$ & [3], this work \\
\hline
\end{tabular*}
\label{param2}
\end{table}

\subsection{Surface Images}\label{images}
\begin{figure*}
\includegraphics[width=1.0\textwidth]{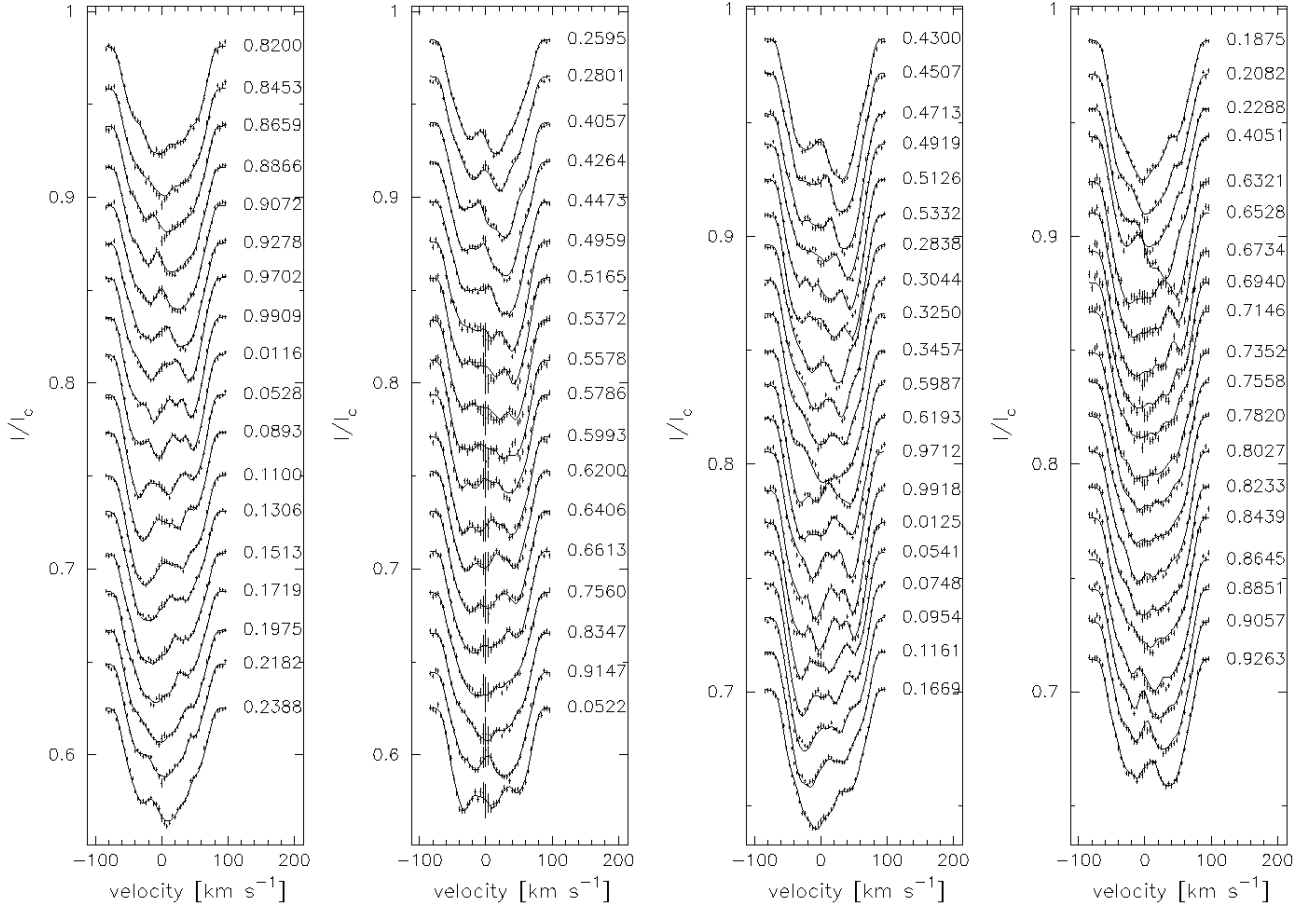}
\caption{Fits (lines) and deconvolved profiles (symbols) for TWA 6 during Feb 2006. The fits for 11--13th are shown in the two boxes on the left, and those for the 17--20th are in the two on the right. The numbers on the right of the plots are the rotational phase at time of observation. The error bars for the data are drawn. The larger error bars in the last 11 profiles in the second box are due to the removal of a feature due to Moonlight in the spectra.  }
\label{fit1}
\end{figure*}

Fig. \ref{fit1} shows the fits to the profiles for both datasets. For the first dataset we achieve a \chisq~of 1.7, in the second dataset it is 2.6. A reduced \chisq~of 1.0 is not achieved for either dataset. The errors estimated in LSD using photon noise statistics are often underestimated in the case of unpolarised spectra \citep{jf97b}. No scaling of errors was carried out in this work, so \chisq~values greater than 1 are expected.  
The normalised residuals of the fits compared to the data were studied and there are no obvious systematic effects increasing the \chisq~or introducing artefacts into the maps.
The distortions in the profiles are very large, suggesting the star has high spot filling factors and sizable temperature differences on the surface. This may explain the previous identification of the TWA 6 as a binary star in \cite{jaya06}, as the presence of a large distortion at the line--centre may cause it to be mistaken for a double-lined binary.

\begin{figure}
\includegraphics[width=0.48\textwidth]{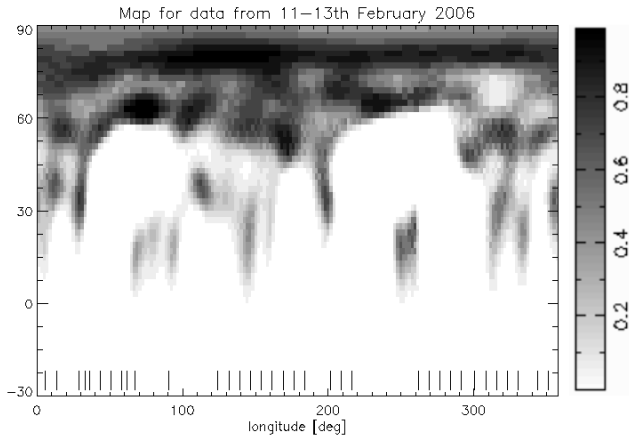}
\caption{Doppler image produced from the data from 11--13th Feb 2006, shown as a Mercator projection. White represents a zero spot--filling factor and a temperature of 4000K, black represents complete spot--filling and a temperature of 3300K. The greyscale is shown on the right hand side, and latitude on the left. Tickmarks along the bottom edge of the map show the phase coverage of the data.The average filling factor $f$ is 9.5\%.}
\label{map1}
\end{figure}

\begin{figure}
\includegraphics[width=0.48\textwidth]{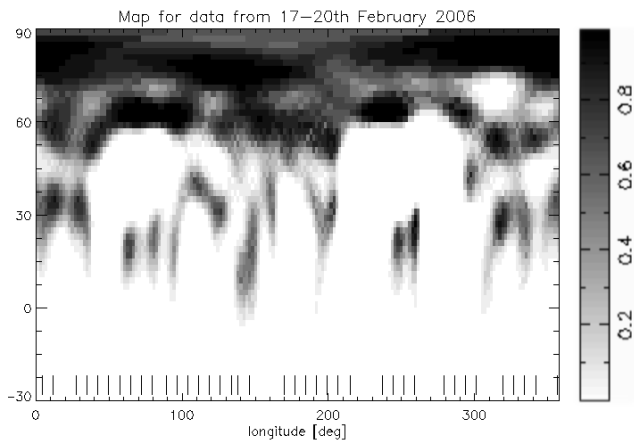}
\caption{As Fig. \ref{map1} for the second set of data (from 17--20th Feb 2006). The average filling factor is 10.7\%.}
\label{map2}
\end{figure}

\begin{figure}
\includegraphics[width=0.48\textwidth]{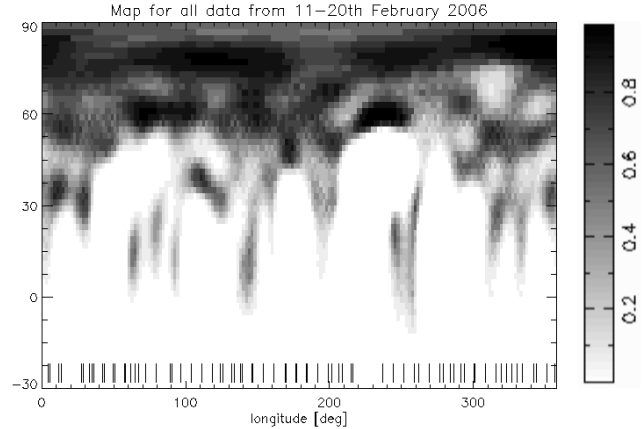}
\caption{As Fig. \ref{map1} for the combination of both data sets. The average filling factor is 9.5\%. }
\label{maptot}
\end{figure}

\begin{figure}
\includegraphics[width=0.48\textwidth]{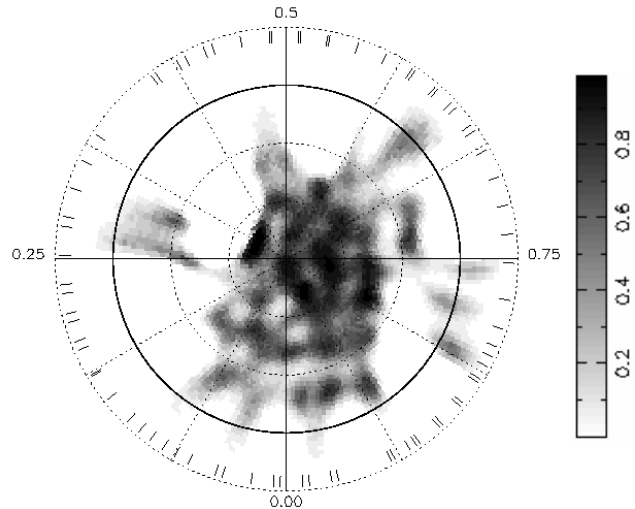}
\caption{A polar projection of the Doppler map for the full dataset for 11-20th February 2006. Dotted circles are at latitude intervals of $30^{\circ}$, the solid circle is at the equator. Longitude intervals of $30^{\circ}$ are also shown . The numbers around the edge show phases from 0--1, corresponding to those given in Fig \ref{fit1}. Tickmarks are placed at phases where observations were made. }
\label{polar}
\end{figure}

The maps given by these fits are shown in Figs. \ref{map1} and \ref{map2}. In all maps white represents a zero spot--filling factor and a temperature of 4000K, black represents complete spot--filling and a temperature of 3300K. Both maps show a polar spot and features extending out of it down to lower latitudes and some isolated spots near the equator. The features are consistent between the two maps, and both are consistent with Fig. \ref{maptot} which was obtained by fitting both datasets simultaneously. The structure of the polar spot can be seen more clearly in Fig. \ref{polar}, which shows the star in a flattened polar projection.

High latitude and polar spots are not observed on the Sun, but the presence of a polar spot is common to other young K stars e.g. LO Peg, \citep{barnes05b} and V410 Tau, \citep{v410a}.  In rapidly rotating stars polar spots have been explained by Coriolis forces causing magnetic flux tubes to rise almost parallel to the rotational axis \citep{schu92}. Magnetohydrodynamical calculations in \cite{caligari94} confirmed that for stars with high rotational velocities and deep convection zones high spot emergence latitudes are expected. 

\cite{gran2000} carried out simulations of flux tube emergence in stars with a variety of rotation rates and masses. According to these models, for a TTS with a mass of 0.7 \mdot~ and an angular velocity 50 times that of the Sun, the predominate flux tube emergence should be within a latitude range $60-90^{\circ}$. The lower latitude spots are not predicted. In the model for rapidly rotating stars ($\Omega \geq 10~\Omega_{\odot}$) with very small radiative cores the Coriolis force is expected to overwhelm the buoyancy forces. The flux tubes detach from the overshoot layer altogether and the resulting flux rings rise close to the pole, giving rise to little spot emergence below $60^{\circ}$. TWA 6 has spot emergence at all latitudes above the equator, and the spot distribution looks more like that for a 1.0 \mdot~star. However given the uncertainty in its mass this is not unreasonable.


\subsection{Photometric Lightcurves}\label{phot}

The Doppler images produced in section \ref{images} were used to produce monochromatic synthetic lightcurves centred at B,V and I. The results were double peaked profiles with the secondary peak being about one-eighth the size of the primary. Photometry is often used as a constraint on spot sizes, e.g \cite {vogt99}, however contemporaneous  photometry of TWA 6 does not exist. We can qualitatively compare our lightcurves with the photometry in \cite{lawson05}, taken in March 2000. In doing this we are not intending to directly compare the lightcurves, as several years elapsed between the two sets of measurements, only to check that our spot maps are plausible in light of previous results.

 As the central wavelength of our LSD profiles is $\sim 5500\rm \AA$ comparison should be made with the V band photometry. The amplitude of the variability, 0.1 magnitudes is around one-fifth of the value measured by \cite{lawson05} ($\Delta \rm{V} = 0.5$). At first sight this suggests that the spot coverage when our measurements were made was rather different to when the photometry in \cite{lawson05} was carried out. It should however be noted that it is rather difficult to reliably recover the amplitude of the variation using spectroscopy alone as spots at low latitudes will be given less weight in Doppler imaging, and the amplitude will be underestimated, see e.g. \cite{unruh95}. 
 
The lightcurve amplitude of  $\Delta \rm{V} = 0.5$ magnitudes is large compared to other WTTS, where amplitudes of $\sim0.1$ are more typical. A check should therefore be made to make sure our spot/photospheric flux differences are realistic. A simple calculation suggests that a spot coverage of approximately 40\% on one hemisphere (i.e. 20\% of the total surface) can recreate the observed brightness change. By using DoTS to produce synthetic images effects like limb darkening can be taken into account. Using our spot temperature of 3300K a lightcurve with an amplitude of 0.5 can be produced with a gaussian spot at $30^{\circ}$ latitude and a mean spot-filling factor of 9\%. Such a spot has a full width at half maximum (FWHM) of $13^{\circ}$. The mean filling--factor in the images in Figs. \ref{map1}, \ref{map2} and \ref{maptot} is 10\%.  This suggests that there was only one major spot group in 2000 whereas in 2006 there were relatively large mid-latitude spots apart from the main group at phase 0.0. Hence it is likely that a combination of underestimation of the variability using Doppler imaging and a change in the spot coverage over the intervening years causes the discrepancy. There is certainly a case for further photometry on this object, as subsequent sections will show.

\subsection{Differential Rotation}\label{diff}

\begin{figure}
\hspace{-2mm}
\includegraphics[width=0.48\textwidth]{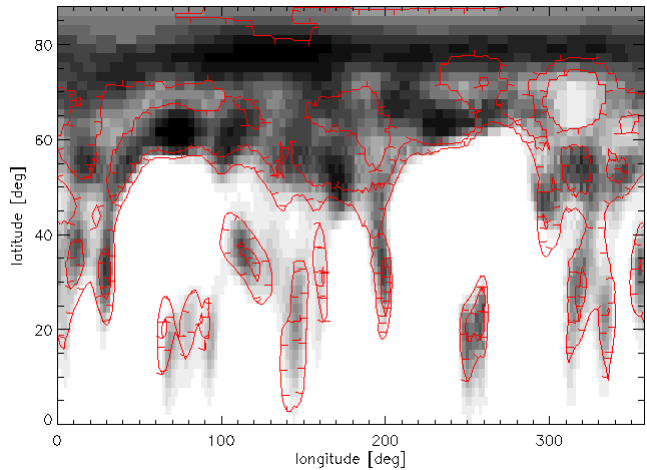}
\caption{The greyscale image from first dataset overlaid with the contours of the second for the northern hemisphere. The two images were reconstructed using similar phases.}
\label{contg}
\end{figure}

Figs. \ref{map1} and \ref{map2} show close agreement between the spots in the two datasets. Fig. \ref{contg} shows the contours of the second map overlaid on the greyscale image of the first. The two images in this case were constructed using similar phases in order to remove differences that may arise due to the different phase coverage of the two datasets. The alignment of the spots suggests that there is no strong differential rotation in TWA 6.  The main differences are in the spots at lower latitudes, where the spots in the greyscale seem slightly larger. However given the difficulties in determining the exact latitudinal extent and weight of low-latitude features some differences are to be expected. There is no strong evidence of the meridional circulation of the type referred to in section \ref{intdiff}.

\begin{figure}
\centering
\includegraphics[width=0.48\textwidth]{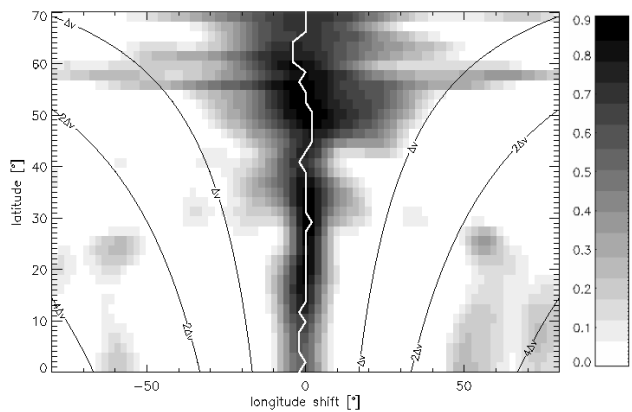}
\caption{Cross correlation of the first map with the second. The solid white line shows the maximum correlation at each latitude. The black curves are a contour map of the velocity represented by each part of the cross-correlation image. $\Delta\rm{v}$ is the velocity resolution of the instrument (6.7\kms).}
\label{cc}
\end{figure}

To quantify the differential rotation more rigorously the two images were cross-correlated at each latitude band with a range of longitude shifts. If differential rotation of the type observed in the Sun was occurring this would be apparent as the band of maximum correlation would curve towards negative longitude shifts at higher latitudes. A solar-type differential rotation can be described by

\begin{equation}
\centering
\Omega(\phi) = \Omega_{0} - \Delta\Omega sin^{2}\phi,
\label{eq:dr}
\end{equation}

\noindent where $\Omega_{0}$ is the angular velocity of the star at the equator, $\phi$ is the latitude and $\Delta\Omega$ is the difference in angular velocities between the equator and the poles. Fig. \ref{cc} plots the value of the cross correlation in a latitude range $0-70^{\circ}$. There is no apparent curvature in the position of the maximum correlation. An upper limit on $\Delta\Omega$ can be calculated by fitting gaussians to each latitude band in the cross correlation. At each $10^{\circ}$ latitude band between $0-70^{\circ}$ the mean position of the gaussian peak is always within one pixel of $0^{\circ}$ (one pixel corresponds to $2^{\circ}$). At higher latitudes it becomes more difficult to cross-correlate the images due to the large polar spot and lower velocity ranges at those latitudes. At lower latitudes the width of the gaussian can provide an upper limit to the differential rotation. The mean width of the gaussians between $0-50^{\circ}$ (where most of the isolated spots are) gives an upper limit on $\Delta\Omega$ of $0.04~ \rm{rad~ day^{-1}}$.

Carrying out a similar cross-correlation, but this time comparing longitude bands between the two images, allows us to investigate whether the star exhibits differential rotation along the meridional direction. As before, a gaussian was fitted to each band in the correlation image (not shown), the mean FWHM of these gaussians gives an upper limit on this form of differential rotation of $0.02~ \rm{rad~ day^{-1}}$. 

\begin{figure}
\includegraphics[width=0.5\textwidth]{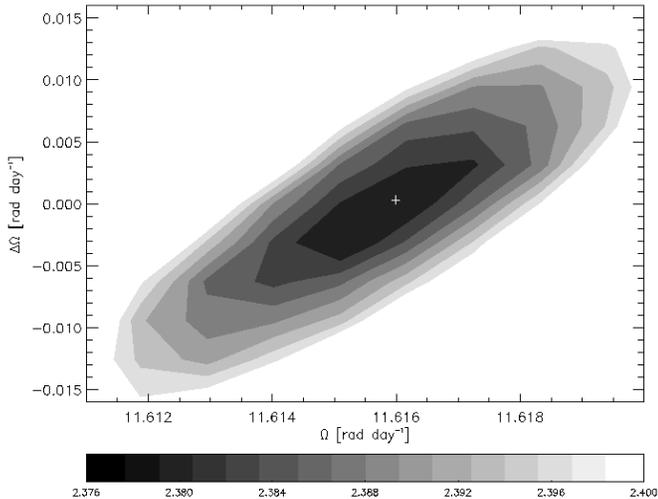}
\caption{A plot of the \chisq~value for a variety of angular speeds and differential rotation values, where $\Omega_{0}$ and $\Delta\Omega$ are as in equation \ref{eq:dr}. The \chisq~values are given in the bar at the bottom. The minimum \chisq, at the `+' in the plot, is at $\Omega = 11.616~ \rm{rad~day}^{-1}$ (corresponding to a period of 0.54091 d) and $\Delta\Omega = 0.0~\rm{rad~day}^{-1}$. The smallest contour is at a \chisq~of 2.376, and approximately corresponds to the $3\sigma$ confidence level. }
\label{drp}
\end{figure}

As a further constraint on the differential rotation DoTS was used to explore the parameter space to find the values which give the minimum \chisq. The relevant parameters which should be varied can be found in equation \ref{eq:dr}. The differential rotation of the star can be parameterised using $\Delta\Omega$, but $\Omega_{0}$ must be varied simultaneously, equivalent to varying the period and differential rotation. The result of this is shown in Fig. \ref{drp}, which is a two-dimensional plot of \chisq~against differential rotation and the angular velocity. We find that the minimum \chisq~is at period = 0.54091 day and differential rotation = $0.00~\rm{rad~day^{-1}}$, further evidence that the star does not have differential rotation.

By drawing contours of $\Delta\chi^{2}$ confidence levels can be placed on the values of $\Omega_{0}$ and $\Delta\Omega$ obtained. When using \chisq~minimization to find a parameter value there is a 67\% probability of the true value lying in the range where $\Delta\chi^{2} \leq 1$ \citep{numrec}. The $1\sigma$ confidence level therefore lies at $\Delta\chi^{2} = 1$.  As our minimum $\chi^{2}$ is more than 1.0 all the $\chi^{2}$ values must be divided by the minimum $\chi^{2}$. The surface is then multiplied by the number of degrees of freedom in the fit, i.e. the number of spectral data points used to create the Doppler image.  The $1\sigma$ confidence level indicates that differential rotation lies within the range $\pm0.003~\rm{rad~day^{-1}}$.

\section{Balmer Line Analysis}

Balmer lines are a useful probe of the circumstellar environment of TTS. H$\rm \alpha$ emission is formed in the stellar chromosphere, disc and wind when hydrogen atoms are ionised by photons from the stellar surface. When a prominence lies in front of the stellar disc from our point of view most of the Balmer line photons will scattered out of our line-of-sight, so prominences in front of the stellar disc appear in absorption. When the prominence is off the stellar disc the photons will be scattered into our line-of-sight and the prominence appears in emission. 

\begin{figure}
\centering
\includegraphics[width=0.45\textwidth]{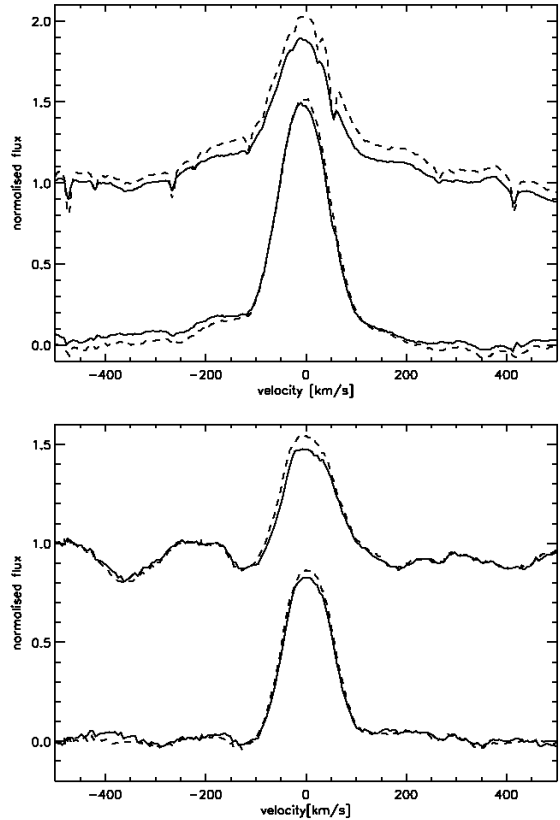} 
\caption{Mean H$\rm \alpha$ profiles (top) and H$\rm \beta$ (bottom). In each plot the upper set of lines shows the normalised mean profile for the first dataset (solid line) and second dataset (dashed line). The lower set of lines are the mean profile with a rotationally broadened absorption profile subtracted off. The template spectrum was rotationally broadened to the \vsini~ of TWA 6 (72\kms), shifted to an appropriate radial velocity, scaled and subtracted. The telluric lines have been divided out. For H$\rm \alpha$ the unsubtracted and subtracted lines have an equivalent width of $5\rm{\AA}$ and $4.2\rm{\AA}$ respectively, while for H$\rm \beta$ it is $1.2\rm{\AA}$ and $1.0\rm{\AA}$.  }
\label{habmean}
\end{figure}

Fig. \ref{habmean} shows the mean H$\rm \alpha$ and H$\rm \beta$ profile shapes for the spectra in the first and second datasets. The raw profiles are shown on top and underneath a broadened absorption spectrum has been subtracted off. The absorption spectrum was created by broadening the normalised template K7 spectrum to a \vsini~ of 72\kms. The telluric lines have been removed. 
 
Both H$\rm \alpha$ and H$\rm \beta$ profiles have a narrow component with a FWHM of 110 \kms and 113 \kms~ respectively. This in good agreement with the expected FWHM of a rotationally broadened spectral line for a star with a \vsini~ of 72\kms, which is 112\kms~ \citep{gray92} if a limb darkening of 0.88 is assumed \citep{limbd}. The subtracted H$\rm \alpha$ profiles have additional broad wings. These are more prominent on the blue side. The subtracted H$\rm \beta$ profiles by contrast appear to only have a narrow component. 

\begin{figure}
\centering
\includegraphics[width=0.45\textwidth]{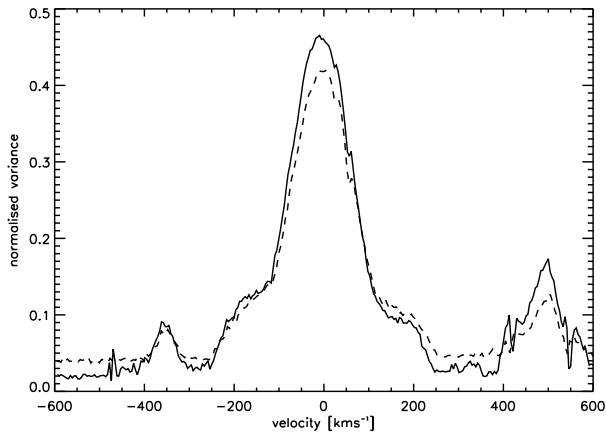}
\caption{The normalised variance profiles for H$\rm \alpha$. The solid line is the variance profiles for the first dataset and the dashed line is for the second dataset. This shows variance out to $\pm$250\kms. The small peak at $\sim -350$\kms~ is due to the presence of an absorption line.}
\label{var}
\end{figure}

The presence of a broad component on the H$\rm \alpha$ emission is supported by the calculation of the variance profile, shown in Fig. \ref{var}. As described in \cite{johns95} this is is calculated using 

\begin{equation}
V_{\lambda} = \Big[\frac{\sum_{i=1}^n(I_{\lambda,i}-\overline{I_{\lambda}})^{2}}{(n-1)}\Big]^{1/2}.
\label{eq:var}
\end{equation}

Fig. \ref{var} shows that the H$\rm \alpha$ emission varies out to $\pm250$\kms, well beyond the \vsini~ of 72 \kms. A similar calculation was carried out for H$\rm \beta$ but this showed negligible variability off the stellar disc and has not been shown here. 

\begin{figure}
\includegraphics[width=0.48\textwidth]{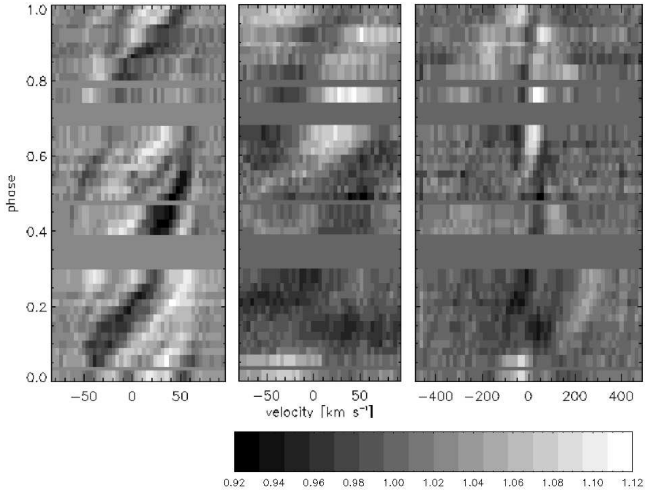}
\caption{Greyscale images for the first set of data (covering 11 -- 13th Feb 2006). From left to right, the normalised deconvolved profiles from Fig \ref{grey}; the normalised H$\rm \alpha$ at velocity range $\pm 90 \rm{km~s}^{-1}$ (bin width 4.2\kms) and the normalised H$\rm \alpha$ at velocity range $\pm 500 \rm{km~s}^{-1}$ (bin widths 17\kms). The scale of the centre and right images is given by the bar at the bottom, while the scale of the left image is as in Fig. \ref{grey}. The similarity between the left and central images should be noted, the bright regions correspond, particularly from phase 0.4 -- 1.0. In the right-hand image there is a sinusoidal feature with an amplitude of 285\kms, see also Fig, \ref{hasin}. }
\label{greyha}
\end{figure}

\begin{figure}
\includegraphics[width=0.48\textwidth]{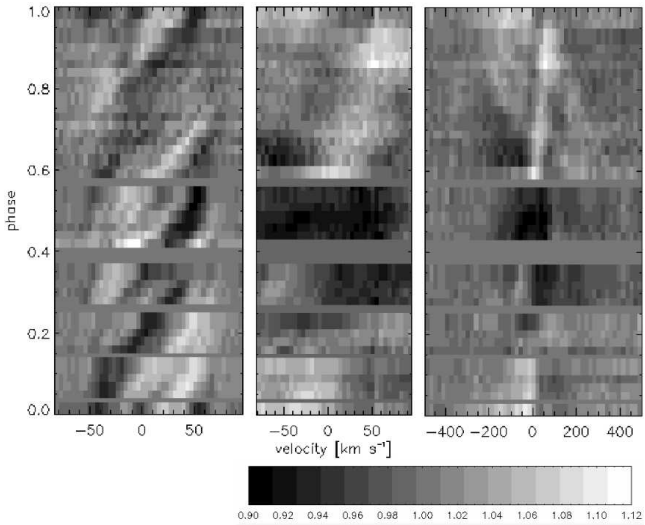}
\caption{As Fig. \ref{greyha} for the second set of data, covering 17--20th Feb 2006. The scale bar refers to the centre and right images, the left image is as in Fig. \ref{grey}. }
\label{greyha2}
\end{figure}

Figs. \ref{greyha} and \ref{greyha2} are greyscale images of H$\rm \alpha$ divided by a mean
profile, at narrow and wide velocity scales, shown with the LSD profiles for comparison. In the left and centre images of Fig \ref{greyha} there is a strong similarity between the LSD and H$\rm \alpha$ profiles. The bright regions coincide in these images, suggesting that the spots on the photosphere correspond with active H$\rm \alpha$ regions. The equivalent H$\rm \beta$ image has similar features, although it is noisier and has not been shown. The SNR in H$\rm \alpha$ is lower than for the LSD as the LSD profiles have been produced using about 2000 lines; we are thus not able to resolve fine detail in the H$\rm \alpha$ images that we can see in the LSD image.

The correspondence is less clear between phases 0 -- 0.4 in Fig. \ref{greyha} than in phases 0.4 -- 1.0. There is a dark feature at around 50 \kms~ in the H$\rm \alpha$ which is making the contrast more difficult to see. In Fig. \ref{greyha2} there is also similarity, between phases 0 -- 0.4 some of the bright regions coincide. At phases 0.4 -- 0.6 in the H$\rm \alpha$ there are dark features masking any possible correspondence but there is some correspondence between phases 0.6 -- 1.0.

\begin{figure}
\includegraphics[width=0.48\textwidth]{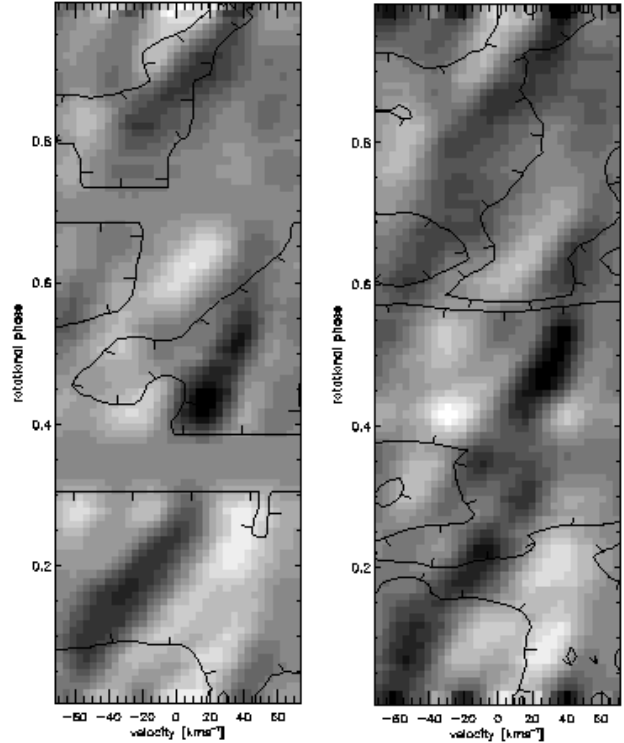}
\caption{LSD profiles shown in greyscale, overlaid with the contours of the H$\rm \alpha$ profiles, for 11--13th Feb (left) and 17--20th Feb 2006 (right). The LSD and H$\rm \alpha$ profiles are as in Fig. \ref{greyha} and \ref{greyha2} but they have been smoothed.}
\label{deccont}
\end{figure}

Fig. \ref{deccont} shows the LSD image with the contours of the H$\rm \alpha$ on top. Both images have been smoothed. The tick marks show the direction of decreasing intensity. The straight horizontal edges in the contour plot are where the phase coverage was incomplete. The alignment between the bright regions is apparent in the left hand image for phase 0.4 -- 1.0. In the right hand image alignment appears to be present between phases 0.6 -- 1.0 only. Note that we have assumed Sun-like rotation, so a phase of 0.4 corresponds to a longitude of $216^{\circ}$. Hence the increased H$\rm \alpha$ emission appears to associated with the group of spots between $60-220^{\circ}$. 

The behaviour implied by Fig \ref{greyha} is similar to the Sun but has not often been observed in other stars. While low-lying plages have been inferred previously, e.g. \cite{jf97b}, \cite{fern04}, we believe that this is the first observation of such a clear alignment of phases and velocities, indicating that the plages are overlying the spots. This is in contrast to observations of earlier type stars such as AB Dor and Speedy Mic where the dominant signal is from higher-lying prominences (known as `slingshot prominences'). 

\begin{figure}
\includegraphics[width=.48\textwidth]{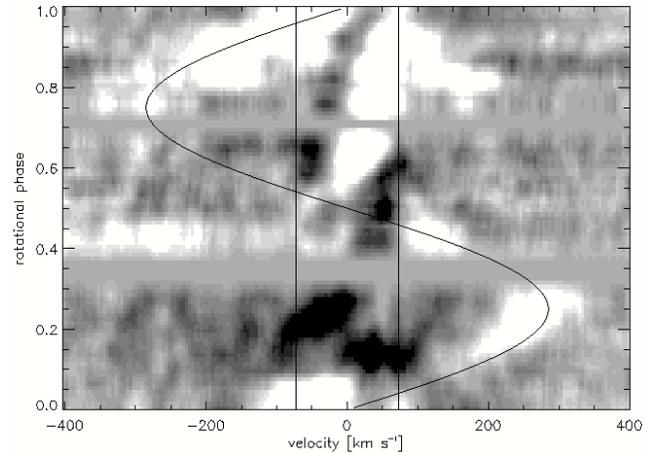}
\caption{H$\rm \alpha$ variability overplotted with a sinusoid indicating the position of a slingshot prominence. The maximum velocity is 285\kms. The vertical lines are at $\pm v \sin i$, i.e. $\pm$ 72\kms. }
\label{hasin}
\end{figure}

Also shown on the right-hand sides Fig \ref{greyha} and \ref{greyha2} is H$\rm \alpha$ variability within a velocity range of $\pm500$ \kms. There are features well outside the \vsini~of the star, suggesting that slingshot prominences can be seen in emission off the stellar disc. In particular there is a bright emission feature in Fig \ref{greyha} that is can be fitted with a sinusoid, shown in Fig. \ref{hasin}. In this figure the image has been smoothed and the contrast has been enhanced, and a sinusoid has been overplotted. The maximum velocity of this sinusoid is 285\kms. If we assume the prominence is corotating with the star then it is lying at a height of about $4R_{*}$.  The star has a corotation radius of $2.4R_{*}$. \cite{jardine05} show that a star of this corotation radius can support prominences out to $4.8R_{*}$. The model assumes that the prominences are embedded in the wind rather than an extended corona. Slingshot prominences have not been detected before in T Tauri stars -- their presence suggests that the star can support stable structures beyond corotation.

Other explanations for this large velocity range should also be considered. For example, there may be infalling gas, close to the surface but with a maximum velocity component along our line of sight of around 285\kms. The observed velocity will vary sinusoidally as the star rotates. Unless the gas is falling directly onto the equator the magnitude of the velocity would not be the same at phases separated by 0.5, as they appear to be here. However if it was on the equator we would not see it for part of the rotation as the inclination of the star would cause it to go out of sight. 

The spectra plotted in Fig. \ref{greyha} were taken over 3 nights, corresponding to about 5.5 rotational periods. All the spectra used in Fig. \ref{greyha} were taken on either the first or third day of observations. Most taken on the third day lie in the phase range 0.5 -- 0.7 where the brightening is less obvious, which implies that the prominence was only present on the first night. However there are two spectra from the third night in the phase range 0.7 -- 1.0 which do support the presence of a prominence, in particular the line at phase $\sim 0.75$ shows a brightening at velocities $\sim 300$ \kms. The prominence may be less apparent at phase 0.5 -- 0.7 because it is on the far side of the star at that time and is obscured by circumstellar material. 

In Fig. \ref{greyha2} there are features outside \vsini~ that may be evidence of more slingshot prominences, particularly between phases 0.6 -- 1.0. However they do not seem to be present at other phases and there is not enough information to justify identifying them as prominences. Also, as they are not at the same phases as the prominence in Fig. \ref{greyha} it is clear that this prominence has not survived the 5 days between the two sets of observations, suggesting lifetimes of the order of a few days. 

\section{Conclusions}\label{con}

Spectra of TWA 6 were taken at AAT/UCLES during two epochs separated by approximately nine rotational periods. Using these spectra and Doppler imaging two maps of its spot distribution were produced. The spot distribution of TWA 6 is qualitatively similar to other T Tauri stars, e.g LO Peg, \citep{barnes05b} and V410 Tau, \citep{v410a}, with the predominate spot emergence at high latitudes and spots all the way to the equator. 

In the Sun photospheric spots are often coincident with `plages', active regions in the chromosphere, usually observed using a H$\rm \alpha$ filter. The H$\rm \alpha$ emission of TWA 6 indicates that the active chromospheric regions are cospatial with the photospheric spots  - similar to the behaviour of the Sun. This supports previous findings that H$\rm \alpha$ emission in young stars increases when spot coverage is larger (\cite{jf97b}, \cite{fern04}) but here we have, for the first time, a strong indication that the plages are overlying the spots. 

TWA 6 appears to support slingshot prominences, the first such observation in a T Tauri star. At least one prominence is observed, lying at a velocity of 285\kms~which corresponds to a radial distance of $4R_{*}$ from the axis. There is evidence that the prominence survives for three days after the initial detection - suggesting that the structure is stable over at least 5.5 rotational periods. The presence of a stable slingshot prominence at a height of $3R_{*}$ from the surface suggests that the star has a large-scale field. Further observations of TWA 6 are justified in order to fully characterise its prominence system, as well as of other stars of similar age and spectral type to determine whether the behaviour set out in this paper is typical. 

The star has no differential rotation (upper limit $0.003~\rm{rad~day^{-1}}$), in line with the previously observed trend that differential rotation decreases with decreasing effective temperature and increasing rotational velocity. Earlier differential rotation measurements have been for older objects, here we have extended the observations to younger stars. The null--measurement is consistent with expectations for the spectral type and angular speed of TWA 6 and indicates that T Tauri stars exhibit solid--body rotation.  
\section*{Acknowledgments}

Thanks to D. Mortlock and N. Dunstone for useful discussions; the staff at the AAT for their help during the observing run and the anonymous referee for helpful comments. The Vienna Atomic Line Database provided the spectral line lists used, and the data reduction was carried out using Starlink software.  M. Skelly would like to acknowledge the financial support of STFC. WAL acknowledges financial support from UNSW@ADFA Faculty Research Grants.

\label{lastpage}
\end{document}